\documentclass{PoS}
\usepackage{color,graphicx}
\usepackage{subfigure}
\usepackage{amsmath}

\title{Partial-Wave Analysis of Centrally Produced Two-Pseudoscalar Final States in \textit{pp} Reactions at COMPASS}

\ShortTitle{Partial-Wave Analysis of Centrally Produced Two-Pseudoscalar Final States in \textit{pp} Reactions at COMPASS}

\author{\speaker{Alexander Austregesilo}\thanks{The author acknowledges financial support by the German Bundesministerium f\"ur Bildung und Forschung (BMBF), by the Maier-Leibnitz-Laboratorium der LMU und TU M\"unchen, and by the DFG Cluster of Excellence 'Origin and Structure of the Universe'. Participation at the conference was financially supported by the organisers.}~~for the COMPASS Collaboration\\
        Technische Universit\"at M\"unchen\\
        E-mail: \email{aaust@tum.de}}

\abstract{COMPASS is a fixed-target experiment at the CERN SPS which focused on light-quark hadron spectroscopy during the data taking periods in 2008 and 2009. A world-leading data set was collected with a 190\,GeV/$c$ hadron beam impinging on a liquid hydrogen target in order to study, inter alia, the central exclusive production of glueball candidates in the light-meson sector. Especially the double-Pomeron exchange mechanism is well suited for the production of mesons without valence quark content. We select centrally produced systems with two pseudo-scalar mesons in the final state from the COMPASS data set recorded with an incoming proton. The decay of this system is decomposed in terms of partial waves, where particular attention is paid to the inherent mathematical ambiguities of the amplitude analysis. Furthermore, we show that simple parametrisations are able to describe the mass dependence of the fit results with sensible Breit-Wigner parameters.}

\FullConference{XV International Conference on Hadron Spectroscopy \\
		 4-8/11/2013\\
		 Nara, Japan}

\begin{document}

\section{Introduction and Kinematic Selection}
  \label{sec:mot}

\begin{figure}[ht]
    \begin{minipage}{.56\textwidth}
      \begin{center}
        \vspace{.5cm}
        \includegraphics[width=\textwidth]{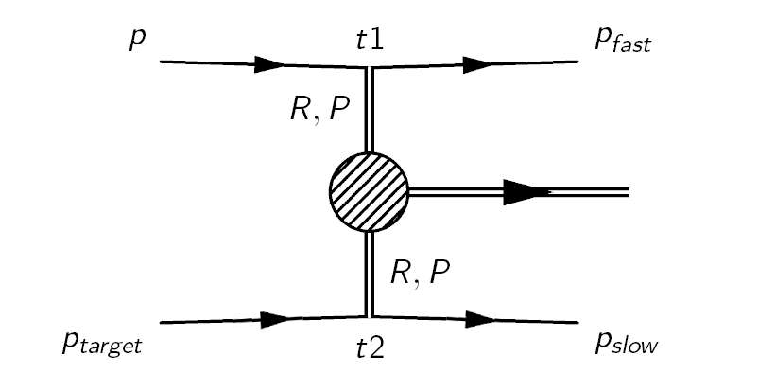}
        \caption{Central production.}
        \label{fig:cp}
      \end{center}
    \end{minipage}
    \begin{minipage}{.44\textwidth}
      \begin{center}
        \includegraphics[width=\textwidth]{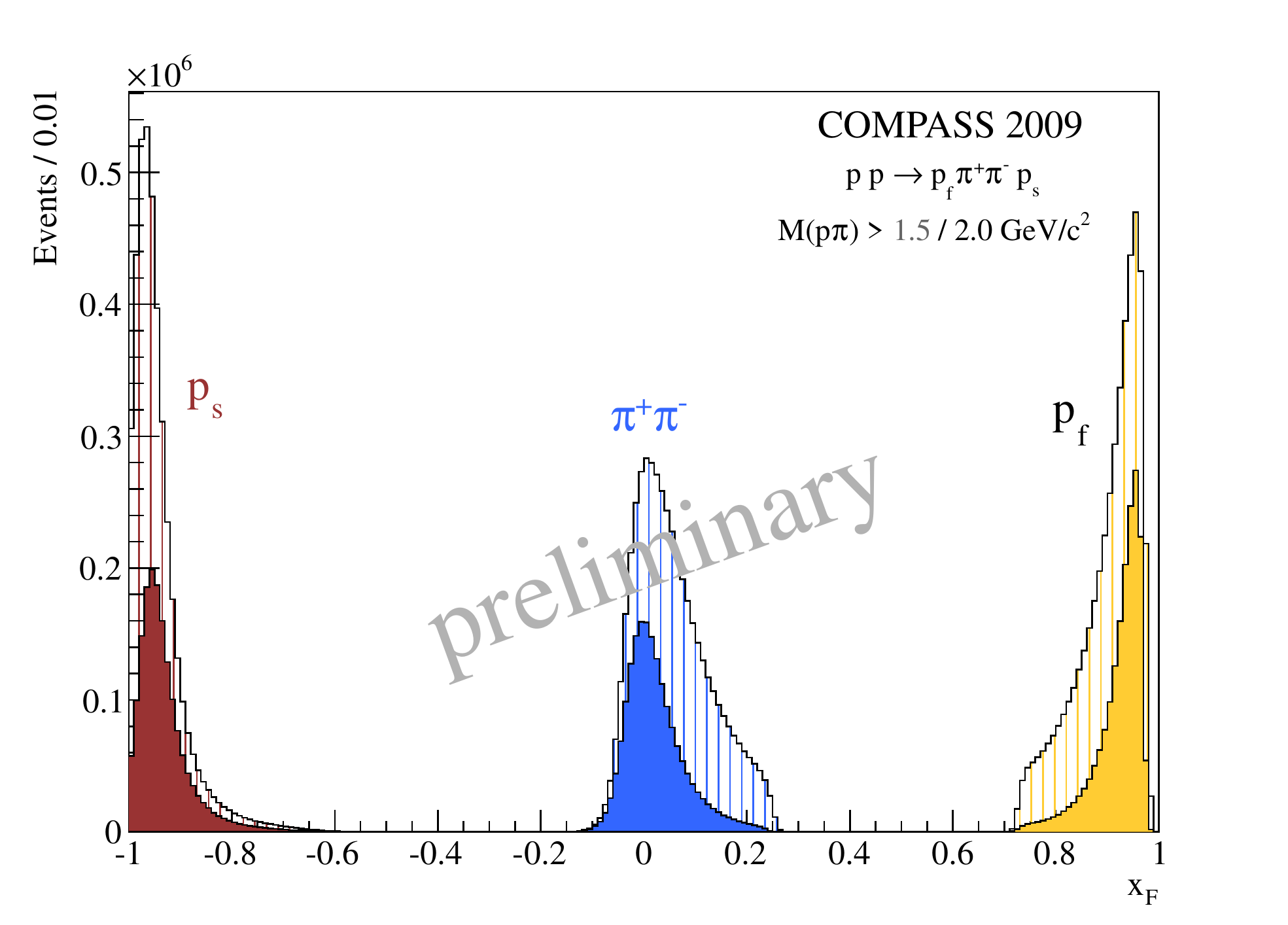}
        \caption[Feynman $x_F$]{Feynman $x_F$ distribution.}
        \label{fig:xf}
      \end{center}
    \end{minipage}
 \end{figure}

One of the goals of the COMPASS experiment~\cite{com07} at CERN SPS is the search for glueballs in the light-quark sector via central exclusive production. This is realised in the fixed-target experiment by the scattering of a $190\,\textrm{GeV}/c$ proton beam on a liquid hydrogen target, where a system of particles is produced at central rapidities (cf. Figure~\ref{fig:cp}).

The trigger on the recoil proton ($p_{\textrm{slow}}$) created a large rapidity gap between the slow proton and the other final-state particles measured in the forward spectrometer. Additional kinematic cuts were used in order to separate the central two-pseudoscalar system from the fast proton $p_{\mathrm{fast}}$. In case of the di-pion system, for example, a cut on the invariant mass combinations $M(p\pi) > 1.5\,\mathrm{GeV}/c^2$ was introduced. The effect of this selection on the Feynman $x_F$ distributions is shown in Figure~\ref{fig:xf}. The $\pi^+\pi^-$ system lies well within $\left \vert x_F \right \vert \le 0.25$ and can therefore be considered as centrally produced.

 \begin{figure}[ht]
   \begin{minipage}{.5\textwidth}
      \begin{center}
        \includegraphics[width=\textwidth]{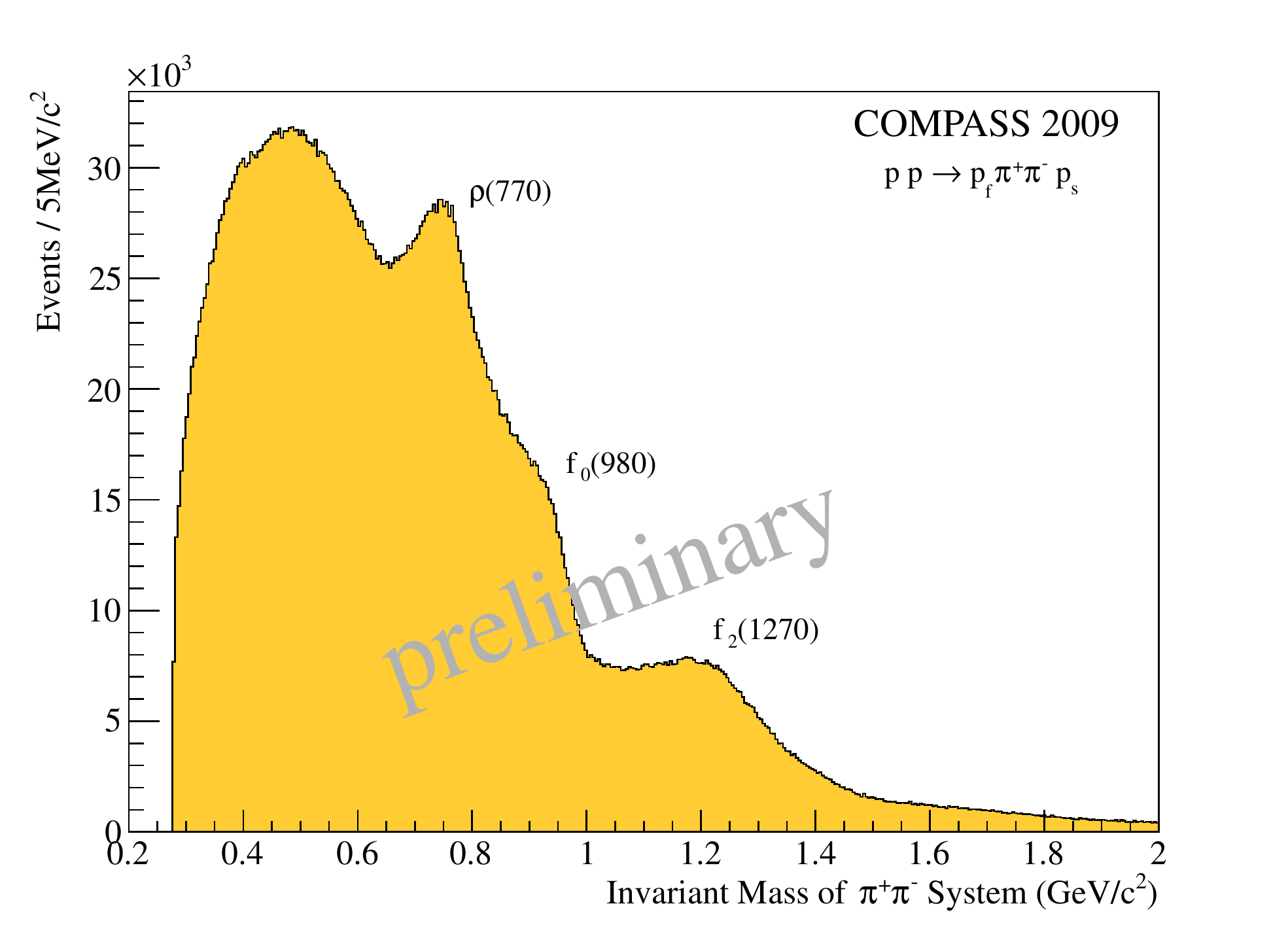}
        \caption[$\pi^+\pi^-$ Invariant Mass]{$\pi^+\pi^-$ system.}
        \label{fig:Mpipi}
      \end{center}
    \end{minipage}
   \begin{minipage}{.5\textwidth}
      \begin{center}
        \includegraphics[width=\textwidth]{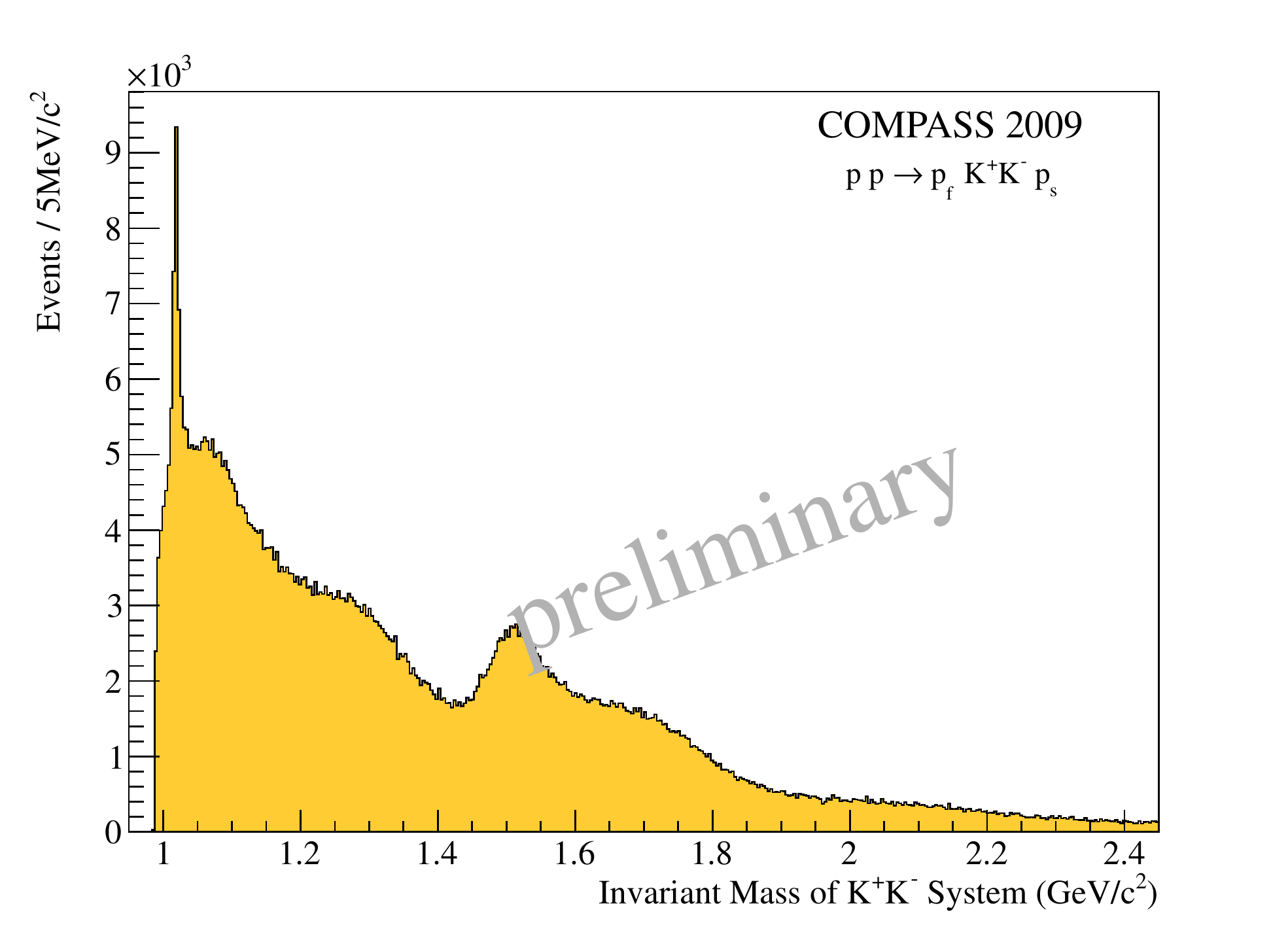}
        \caption[$K^+K^-$ Invariant Mass]{$K^+K^-$ system.}
        \label{fig:MKK}
      \end{center}
    \end{minipage}
\end{figure}

Figure~\ref{fig:Mpipi} shows the invariant mass spectrum of the central $\pi^+\pi^-$ system, where the $\rho$(770), the $f_2$(1270), and the sharp drop in intensity in the vicinity of the $f_0$(980) resonance can be observed as dominant features. Since the $\rho$(770) cannot be produced via double-Pomeron exchange (DPE), contributions from other production mechanisms are evidently non-negligible at $\sqrt s = 19\,\mathrm{GeV}/c^2$. A variety of selection criteria has been used in the past, and several of them have been studied with our data set. Most of them showed a large overlap, but none was able to successfully single out a clean DPE sample. However, the results discussed below exhibit little dependence on this choice.

A rich invariant-mass spectrum is observed for the central $K^+K^-$ system (cf. Figure~\ref{fig:MKK}). In contrast, the only unambiguously identified resonance is the narrow vector meson $\phi$(1020) which, like the $\rho$(770), cannot be produced via DPE. The structures above the $\phi$(1020) will be disentangled by the partial-wave analysis discussed in Section~\ref{sec:massdep}.

  \section{Ambiguities in the Amplitude Analysis of the $\mathbf{\pi\pi}$-Systems}

\begin{floatingfigure}[l]
  \includegraphics[width=4.3cm]{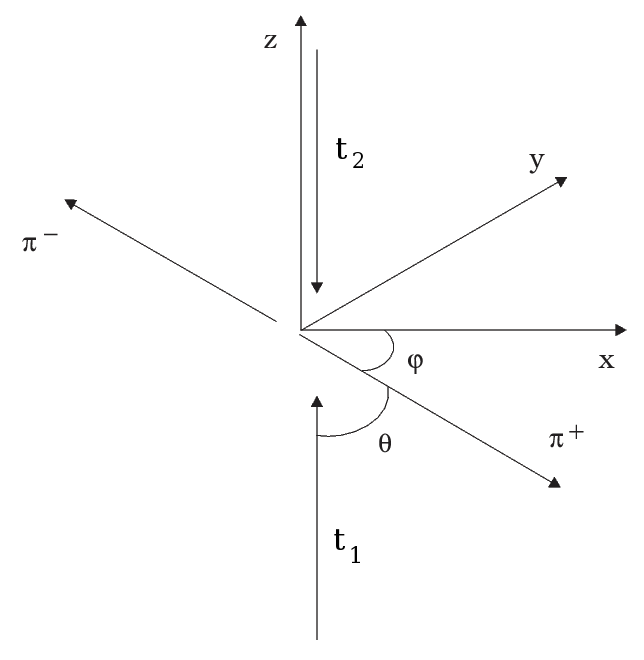}
  \caption{Coordinate system in the $\pi^+\pi^-$ centre-of-mass}
  \label{}
\end{floatingfigure}
The partial-wave analysis has been performed assuming that the central two-pseudoscalar system is produced by the collision of two particles which are emitted by the scattered protons and which form the $z$-axis in the centre-of-mass frame of the $\pi^+\pi^-$ system. The $y$-axis of the right-handed coordinate system is defined by the cross product of the momentum vectors of the two exchange particles in the $pp$ centre-of-mass system. Apart from the invariant mass, the polar and azimuthal angles $\cos\theta$ and  $\phi$ of the negative particle, measured in the two-pseudoscalar centre-of-mass frame relative to the axes defined above, characterise the decay process.
 With complex transition amplitudes, the intensity can be expanded in terms of spherical harmonics. An extended maximum-likelihood fit in $10\,\mathrm{MeV}/c^{2}$ wide mass bins is used to find the amplitudes, such that the acceptance corrected model matches the measured data best. The transition amplitudes are constant over the narrow mass bins, which means no mass-dependence is assumed at this stage.

 \begin{figure}[b]
   \begin{center}
     \subfigure[]{
       \includegraphics[width=.45\textwidth]{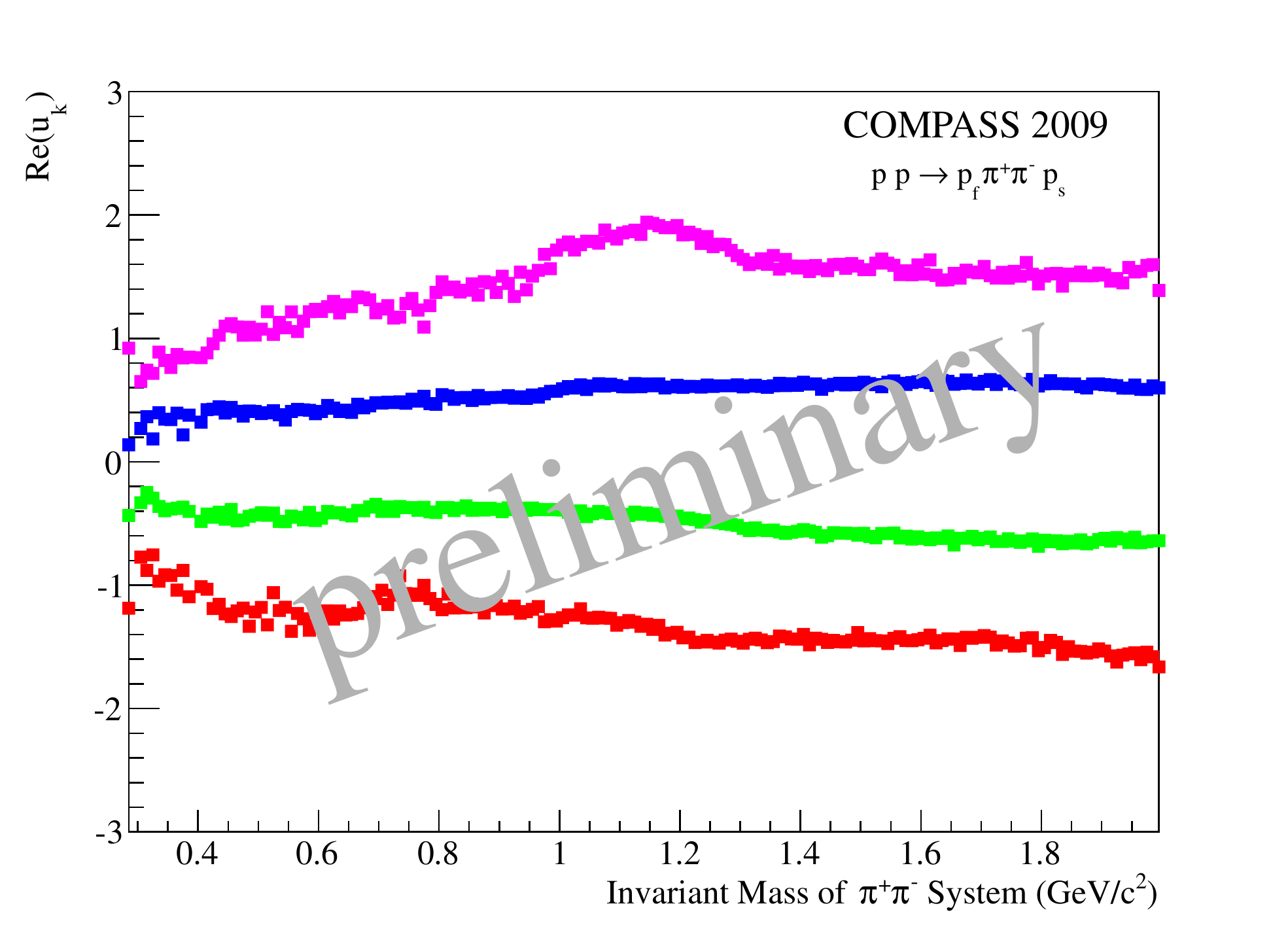}
     }
     \subfigure[]{
       \includegraphics[width=.45\textwidth]{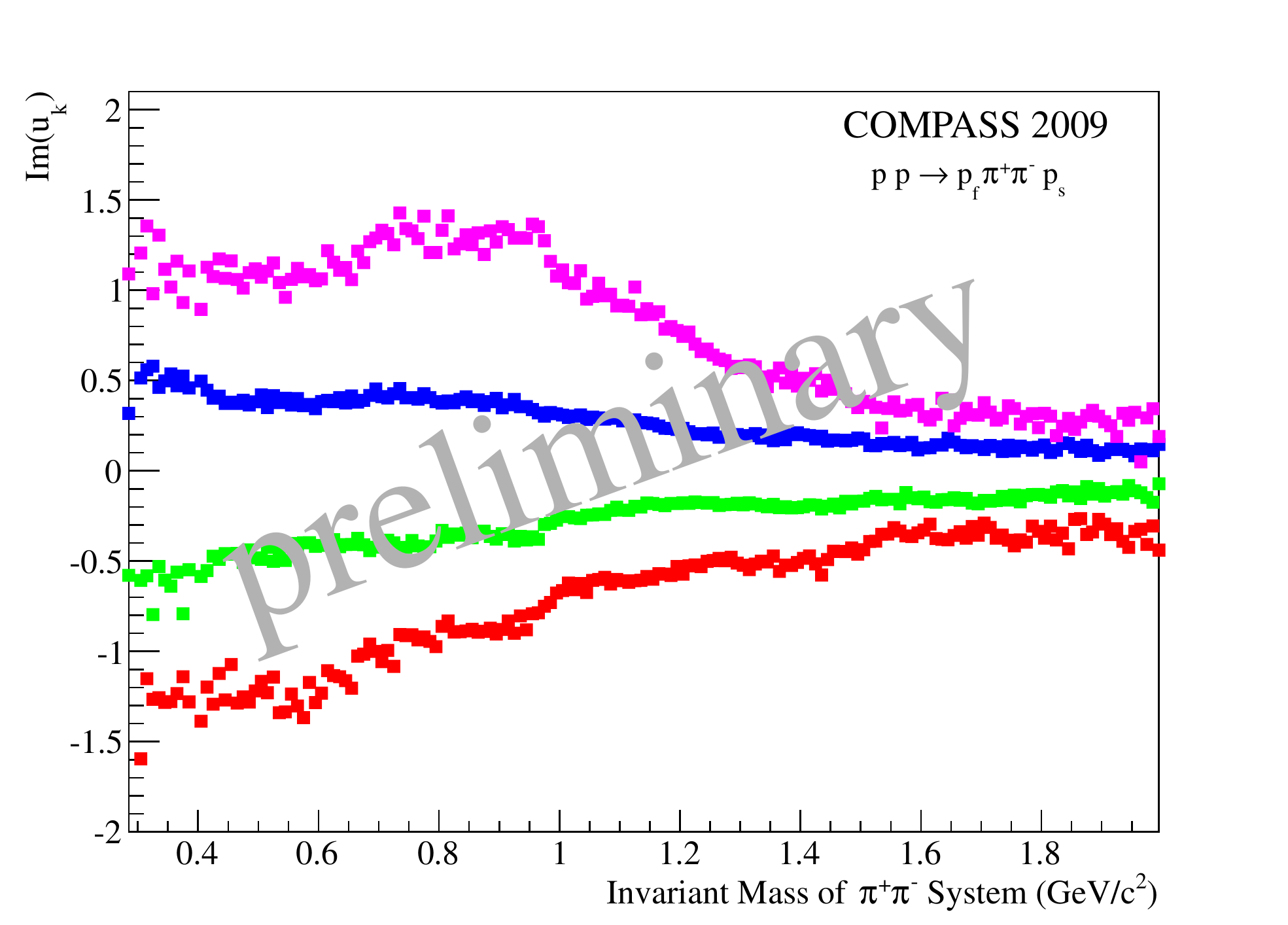}
     }
   \end{center}
   \caption[Barrelet Zeros]{The (a) real and (b) imaginary parts of the polynomial roots as a function of the $\pi^+\pi^-$ mass.}
   \label{fig:roots}
 \end{figure}

 As demonstrated in \cite{sad91}, the system of $S$, $P$ and $D$ waves used in the presented analysis has eight ambiguous solutions. 
 However, the fitted production amplitudes for one single attempt can be used to calculate all eight solutions analytically by complex conjugation of the roots of a $4^{\textrm{th}}$-order polynomial in angular variables~\cite{bar72, chu97}. 
 Figure~\ref{fig:roots} illustrates the real and imaginary parts of these four roots for all mass bins of the $\pi^+\pi^-$-system, sorted by their real part. They are well separated from each other and can be easily linked from mass bin to mass bin. The imaginary parts do not cross zero, hence bifurcation of the solutions does not pose a problem and the eight solutions can be uniquely identified. 

 Differentiation among these mathematically equivalent solutions requires additional input, e.g. the behaviour at threshold or the expected physical content. Complementarily, only waves with even spin are allowed for an equivalent analysis of the $\pi^0\pi^0$ system, which has the advantage of only two ambiguous solutions. A combined analysis of both channels is currently under study.

 \section{Mass-Dependent Parametrisation of the Centrally Produced $\mathbf{K^+K^-}$ System}
 \label{sec:massdep}

 If we apply the same analysis technique to the centrally produced $K^+K^-$ system, the choice of the single physical solution becomes clearer. Only for one solution, the expected dominance of the $S$-wave at threshold is observed. In addition, this solution shows almost no intensity in the $P$-wave above the narrow $\phi$(1020), a fact that supports the assumption of double-Pomeron exchange as the dominant production process. 
For this reason, we limit the analysis to spin-0 and spin-2 contributions in the mass range above $1.05\,\mathrm{GeV}/c^2$. 

\begin{figure}[t]
   \begin{center}
     \includegraphics[width=.85\textwidth]{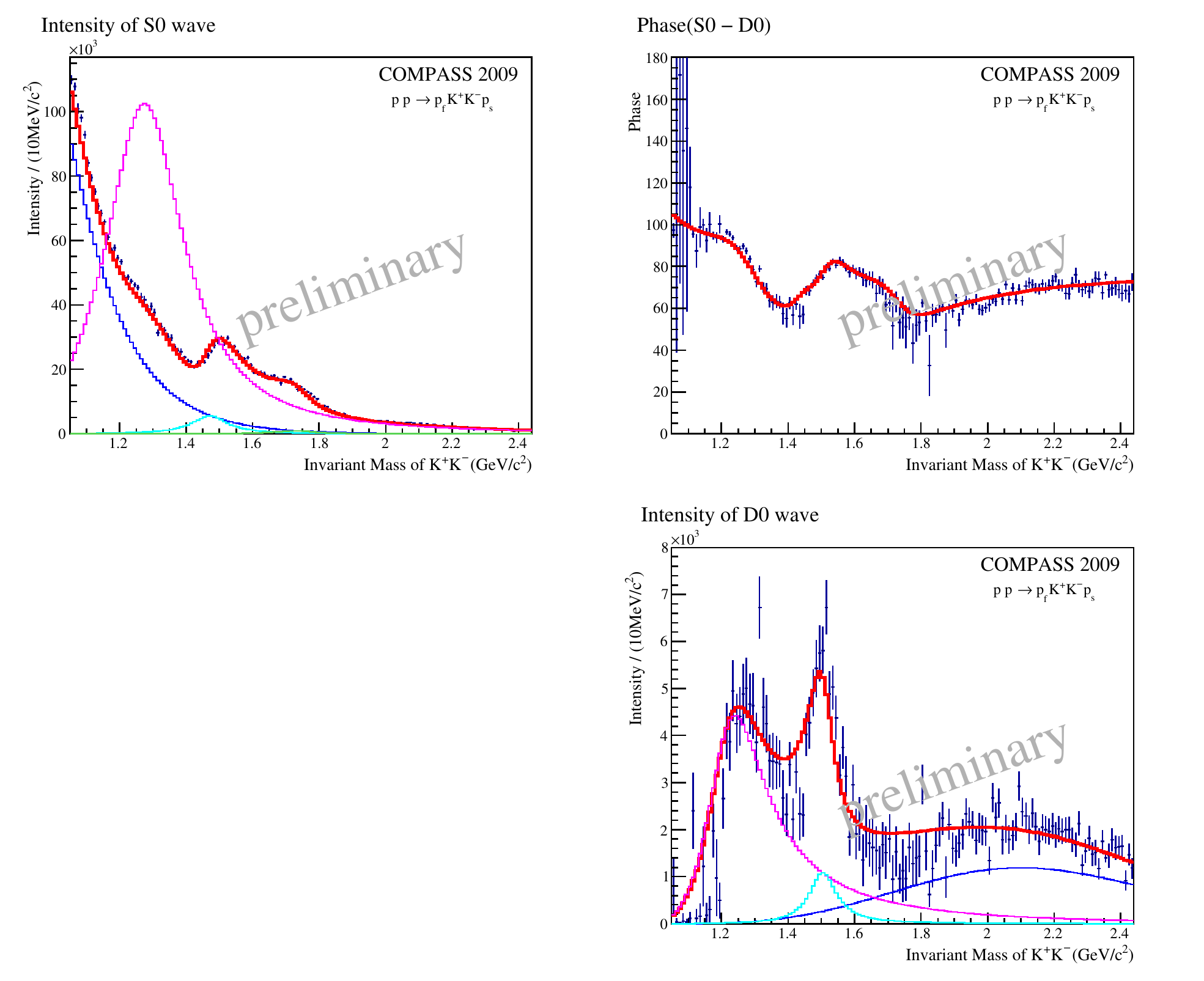}
   \end{center}
   \vspace{-3.cm}
   \begin{minipage}{.48\textwidth}
     \caption[Mass-dependent parametrisation of intensities and phase]{Mass-dependent fit (red curve) of the PWA in mass bins of the $K^+K^-$ system (data points) with non-resonant contribution (dark blue) and Breit-Wigner functions (other colours, see text)}
     \label{fig:fitSD}
   \end{minipage}
 \end{figure}

Starting from these results, the mass dependence of the partial-wave intensities and their interference is parametrised in terms of a physical model. The model parameters are determined by a $\chi^2$-fit to both the real and imaginary parts of the spin-density matrix elements. In this preliminary analysis, we focused only on the two most prominent contributions: the real (anchor) wave $S_0^-$ and the complex-valued $D_0^-$. The resonant contributions are modelled with dynamic-width relativistic Breit-Wigner functions.
Besides the resonance parameters $m_0$ and $\Gamma_0$, we allow for a free complex amplitude for every Breit-Wigner function. A non-resonant contribution has to be introduced in order to account for the threshold enhancement in the $S$-wave as well as for the contributions from other production processes which translate through their angular characteristics into the intensities. This component is modelled by a two-body phase space with the  correct asymptotic behaviour for the angular-momentum barrier and an exponential damping for high masses. The final result of the fit is illustrated in Figure~\ref{fig:fitSD}.

 The two sharp peaks in the $D$-wave intensity were fitted with two well-known resonances. Breit-Wigner functions are used to parametrise the $f_2$(1270) and the $f_2'$(1525) mesons. An additional $f_2$(2150) as suggested by \cite{bar99} was not needed in order to describe the data, the intensity can be attributed to the non-resonant model component. At least three different Breit-Wigner functions 
 were necessary in order to describe the $S$-wave. In addition to the well-established $f_0$(1500) and $f_0$(1710) resonances, a broad $f_0$(1370) had to be included to account for both the intensity as well as the phase with respect to the $D$-wave in this mass region. A strong interference with the non-resonant contribution even required a dominant contribution from this term~(cf.~Figure~\ref{fig:fitSD}). However, the large correlation between the parameters of the $f_2$(1270) in the $D$-wave and the $f_0$(1370) in the $S$-wave may lead to systematic errors in the Breit-Wigner parameters which are still under investigation. In addition, the $f_0$(1370) strength is very sensitive to the parametrisation of the non-resonant component below, and the influence of the $f_0$(980) below the applied mass threshold may also play a role. For this reason, we do not quote the mass and width parameters of the resonant contributions here.


  \section{Outlook}
  \label{sec:DaO}
  
COMPASS is able to select centrally produced two-pseudoscalar final states and describe the main features of the data in terms of partial waves. A more detailed description of the analysis methods can be found in~\cite{aus13}.

The amount of data as well as the sensitivity of the analysis largely exceed earlier studies. The phase relations emerge with unprecedented precision and provide important information ignored by previous analyses~(e.g.~\cite{bar99}). The data show that COMPASS may be able to contribute to the controversial discussion about the resonances in the scalar sector~\cite{och13}. In order to interpret the composition of the super-numerous scalar resonances, a combined analysis of all available final states will be essential. Especially the combination with the corresponding neutral final states $\pi^0\pi^0$, $\eta\eta$, and $K^0_SK^0_S$ can help to resolve remaining ambiguities.

\end{document}